\documentclass[aps,physrev,preprint,groupedaddress]{revtex4-2}
\usepackage{color} %used for font color
\usepackage{multirow}
\usepackage{xcolor} %used for font color
\usepackage{amssymb} %maths
\usepackage{amsmath,bm} %maths
\usepackage{siunitx}
\usepackage[utf8]{inputenc} 
\usepackage{graphicx}
\usepackage{tabularx,setspace,booktabs,boldline,scrextend,hhline,makecell}
\usepackage{epstopdf}
\usepackage{hyperref}
\begin{document}

%\preprint{}

\title{FIRST SYNCHROTRON INJECTION ATTEMPT INTO THE SuperKEKB HER}


\author{N.~Iida\thanks{naoko.iida@kek.jp}, Y.~Funakoshi, H.~Kaji, T.~Kamitani, M.~Kikuchi, K.~Kodama, H.~Koiso, T.~Mimashi, \\G.~Mitsuka, T.~Mori, Y.~Ohnishi, Y.~Seimiya, K.~Shibata, H.~Sugimoto, M.~Tawada, T.~Ueda, R.~Ueki, \\T.~Yoshimoto, High Energy Accelerator Research Organization, Tsukuba, Japan\\ 
M.~Li, Chinese Academy of Sciences, Beijing, China\\ 
K.~Oide, European Organization for Nuclear Research, Geneva, Switzerland}
%\thanks{}
\affiliation{}


\date{\today}

\begin{abstract}
A synchrotron injection scheme for the SuperKEKB electron high-energy ring (HER) was implemented and experimentally evaluated as the first attempt with a top-up injection during beam collisions.
The lattice at the HER injection point was configured to provide a large horizontal dispersion of \,–1.6 m, and the injection beam energy was accordingly set to +0.6\% above the ring energy. 
Since the beam extraction system is located near the injection point, the optics design was constrained to ensure compatibility with its requirements.
A systematic tuning procedure of the injection parameters has been established with turn-by-turn BPMs(TbT-BPMs) in the ring, by which the betatron amplitude of the injected beam was successfully removed. 
Using optimized ring optics, synchrotron injection into the HER was successfully demonstrated, followed by the establishment of stable collisions and the production of luminosity. 
%As the injection repetition rate from 5\,Hz to 12.5\,Hzwas increased, however, an unexpected degradation in injection efficiency was observed, leading to the suspension of the study.  
%These experimental results and some remaining issues will be reported.
\end{abstract}
%\keywords{}

\maketitle

\section{INTRODUCTION}
The peak luminosity of SuperKEKB~\cite{TDR,Ohnishi} is currently 5.28$\times$10$\rm ^{34}/cm^2s$\cite{Luminosity}, while the next target is 1$\times$10$\rm ^{35}/cm^2s$.
To achieve this target, it is essential to increase the injection rate for both electrons and positrons. 
However, at present, improving the injection efficiency remains challenging in both rings~\cite{MengLi, eeFACT2025}.
There are at least two primary reasons for this. 
The first is the growth of the injected beam emittance through the beam transport line (BT) after the injector Linac down to the collider rings. 
For both electron and positron beams, the transverse emittances increase along the BTs by factors of 2 to 5.
The second is the beam loss occurring after injection in the storage ring. 
This loss has several contributing factors, one of which is the horizontal injection oscillation.
The current injection scheme of SuperKEKB employs on-momentum, off-axis injection, called ``betatron injection''(BI). 
%In this scheme, the injected beam undergoes large horizontal oscillations in the ring, leading to various disadvantages. 
This scheme leads to some performance issues, since the injected beam undergoes a transverse oscillation in the ring.
However, by adopting off-momentum, on-axis injection, so-called ``synchrotron injection''(SI)~\cite{LEP, Liu, Aiba}, these issues can be mitigated, as described below.

\subsection{Horizontal Injection Oscillation}
In general, when the injected beam undergoes horizontal oscillations during top-up injection,
%it becomes susceptible to beam–beam kicks from the opposing beam during collisions,
it becomes susceptible to beam–beam kicks from the opposing beam,
which can lead to beam blowup and tail formation.
This is true even for head-on collisions; however, as SuperKEKB employs the nano-beam scheme~\cite{NanoBeam} for collisions, where beams collide with a large horizontal crossing angle.
As a result, if the injected beam has a large horizontal amplitude, it collides at a $s$-shifted location where the vertical beta function is large, and the resulting kick may induce beam loss.
In SI, the injected beam does not have horizontal oscillations, and therefore such beam loss does not occur. In fact, this scheme had already been considered during the design stage of SuperKEKB~\cite{TDR:safety}.

\subsection{Narrow Dynamic Aperture of the HER}
In the HER, the injection efficiency (survival ratio) does not exceed $\sim80$\,\% even 100 turns after injection. 
One of the causes is believed to be that the injected electron beam, having a large vertical emittance and a large horizontal injection oscillation, exceeds beyond the dynamical aperture of the HER. 
%Although the current vertical beta function at the interaction point(IP) is $\beta_y^*=1\,{\rm mm}$, measurements using turn-by-turn beam position monitors (TbT-BPMs) have revealed that the actual vertical dynamical aperture seems smaller than the design value\cite{}.
Furthermore, the dynamical aperture has been reduced due to ``mis-wiring''~\cite{Ohnishi,IPAC2023} of the cancel coils in the superconducting final focus magnet (QCS) during construction, making it even more restrictive for injected beams undergoing large horizontal oscillations. 
In SI, however, the injected beam no longer exhibits horizontal oscillations, and therefore an improvement in injection efficiency is expected. 
Although the injection oscillation is transformed from horizontal to  longitudinal plane, simulations described later indicate that the associated disadvantages are much smaller, since the dispersions at the interaction point(IP) is suppressed.

\subsection{Synchrotron Radiation from Injection Beam}
In the case of BI, the injected beam undergoes horizontal oscillations, and synchrotron radiation generated by kicks from the high-field QCS magnets can hit the beam pipe near the IP, potentially contributing to background in the Belle II detector.

\subsection{Damping Time}
Since the longitudinal damping time in the HER, 2900 turns, is a half of that in the transverse plane, the background affecting the Belle II detector is expected to decay more rapidly.

\section{Optics and Orbits}
Optics solutions for SI were developed and optimized\cite{Oide:2016mkm} using the ring sextupoles to enlarge the dynamic aperture for $\beta_y^*$ values of 81\,mm, 8\,mm, 3\,mm, and 1\,mm at the IP. The resulting dynamic momentum acceptances for $\beta_y^*=1\,$mm are shown in Fig.~\ref{fig:injDA} shown later.
These resulting optics are used in the simulations described below and the actual beam operation for SI. However, the so-optimized optics for BI was only used in the simulations below. 

\begin{figure}[htb!]
   \centering
\includegraphics[width=0.8\columnwidth]{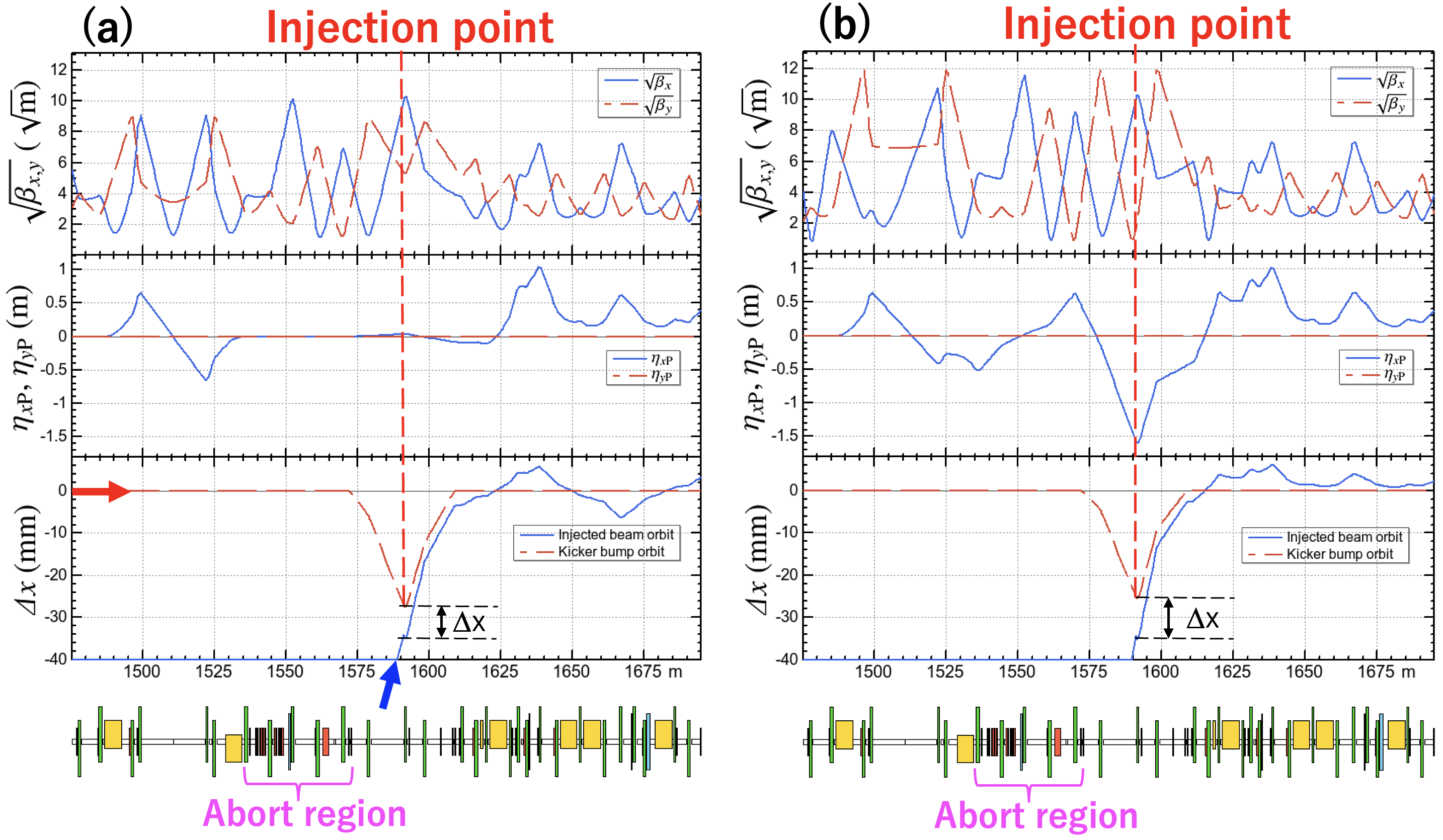}
   \caption{Optics and orbit plots for (a) BI and (b) SI. 
   The top and middle plots show the beta and dispersion functions (blue: horizontal, red: vertical). 
   The bottom plot shows the horizontal orbit: red and blue lines show the stored-beam kicker bump orbit and the injected beam orbit, respectively. 
   The $\varDelta x$ shows the horizontal distance between stored and injected beam at the injection point.   \label{fig:optics}}
\end{figure}

Figure~\ref{fig:optics} shows the optics and orbit in the injection region~\cite{TDR:safety}. 
In conventional BI (a), there is no dispersion in this region, whereas in SI (b), a large horizontal dispersion of -1.6\,m is introduced. 
In (a), the injected beam acquires horizontal oscillation due to the separation between the injected and stored beams ($\Delta x$), which converges via the radiation damping in the ring.
In (b), the injected beam, with 0.6\,\% higher energy than that of the stored beam, follows the dispersive orbit. 
The horizontal offset $\Delta x$ is translated into the longitudinal phase space as the dispersion decreases.
This dynamics in SI has an advantage in the beam--beam lens to that in BI.
%In regions without dispersion like the interaction region, the horizontal orbit offset vanishes.
The beam extraction/abort region is located in the same straight section as the injection region (Fig.~\ref{fig:optics}).
Therefore, the horizontal dispersion region for SI has been created without changing the transfer matrix between the extraction kicker and the beam dump entrance from that in BI.
%A key consideration in the optics is that the beam extraction region shown in Fig.~\ref{fig:optics} is included within the horizontal dispersion region created for SI. 
%It is necessary to keep the transfer matrix between the extraction kicker and the beam dump entrance unchanged from that of BI.
%the titanium extraction window, through which the extracted beam exits to the air, and 

\begin{figure}[!htb]
   \centering
   \includegraphics[width=1\columnwidth]{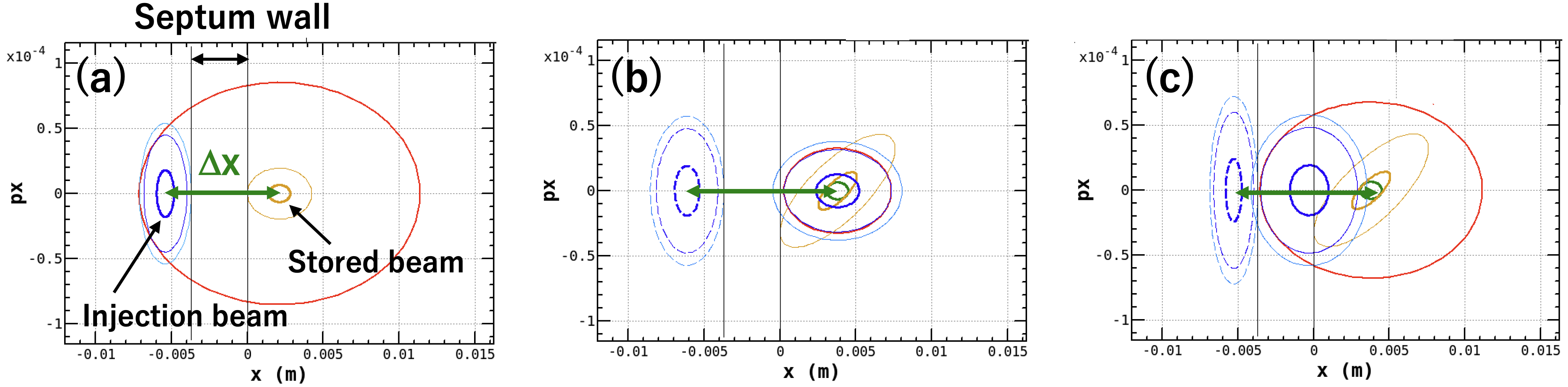}
   \caption{Horizontal phase space at the injection point for (a) betatron, (b) synchrotron, and (c) synchro-beta injection, where the virtual injection point ({\tt PINJAX0}) where $\alpha_x$=0. 
   Pale yellow and blue ellipses indicate the 3$\sigma$ stored beam and 2.5$\sigma$ injected beam, respectively; red shows the action of the injected beam in the ring.   \label{fig:action}}
\end{figure}

Fig.~\ref{fig:action} shows the phase-space distributions of the stored and injected beams at the injection point. 
In BI (a), the stored beam is brought close to the septum wall on the ring side by the kicker bump orbit, while the injected beam approaches the septum from the injection side. 
The offset $\varDelta x$ shown in Fig.~\ref{fig:action} induces a horizontal injection oscillation and persists until it damps. 
In this case, the $\beta_x$ of the injected beam is optimized to minimize the resulting action $J_x$ in the ring, and therefore differs from that of the ring.
In SI in Fig.~\ref{fig:action} (b), the Twiss parameters of the injected beam are set equal to those of the ring, so the beam fully matches the ring ellipse after injection. 
As is clear from Fig.~\ref{fig:action}, the post-injection action (red) is the smallest among the three schemes. 
The elongation of the injected beam in the $P_x$ direction in Fig.~\ref{fig:action} (b) arises because $\eta_{p_x}$ is no longer zero when matched to the ring, leading to a difference $\delta \eta_x'$ from the ring.
In ``synchro-beta injection''(SBI) (c) distributes the injection oscillation between the horizontal and longitudinal directions in which the ring optics are the same as for SI, while the injected beam energy is set to +0.3\,\% of the ring energy in this example.

The parameters for BI, SI, and SBI are summarized in Table~\ref{tab:parameters}.

\begin{table}[!hbt]
    \centering   
    \caption{Injection Parameters. 
    The subscript ``${\rm inj}$'' denotes the values of ring / injection beams at the virtual injection point {\tt PINJAX0} where the $\alpha_x$ is zero. The betatron component of the injection oscillation and the initial momentum offset  are shown as $\varDelta X_\beta$ and $\varDelta p/p_0$, respectively.
    }
 % \begin{minipage}{\columnwidth}
  \small
    \begin{tabular}{lccc}
    \toprule
   & \textbf{BI} & \textbf{SI} & \textbf{SBI} \\
\midrule
$\varepsilon_x$ [nm] &  & 4.6 / 10.2 &  \\
$\sigma_\delta$ [\%] &  & 0.063 / 0.18\textsuperscript{a}  \\
$\beta_{x,{\rm inj}}$ [m] & 108.4 / 31.6 & 108.4 / 63.9 & 108.4 / 27.2 \\
$\alpha_{x,{\rm inj}}$ & 0 / 0 & 0 / 0 & 0 / 0 \\
$\eta_{x,{\rm inj}}$ [m]  & 0 / 0 & -1.66 / 0 & -1.66 / 0 \\
$\eta_{p_x,{\rm inj}}$ & 0 / 0 & -0.02/-0.02 & -0.02 / -0.02 \\
$\varDelta X$ [mm] & -7.52 & -9.91 & -9.07 \\
$\varDelta X_\beta$ [mm] & -7.52 & 0.0 & -4.09 \\
$\varDelta p/p_0$ [\%] & 0.0 & +0.6 & +0.3 \\
$2J_x$ [{\unit \um}] & 0.79 & 0.12 & 0.50 \\
\bottomrule
   \multicolumn{4}{@{}l}{\footnotesize\textsuperscript{a} full width}
   \end{tabular}
%\end{minipage}
   \label{tab:parameters}
\end{table}
%\vspace{-5 mm}

\section{Simulations}
\begin{figure}[!htb]
   \centering
   \includegraphics*[width=1\columnwidth]{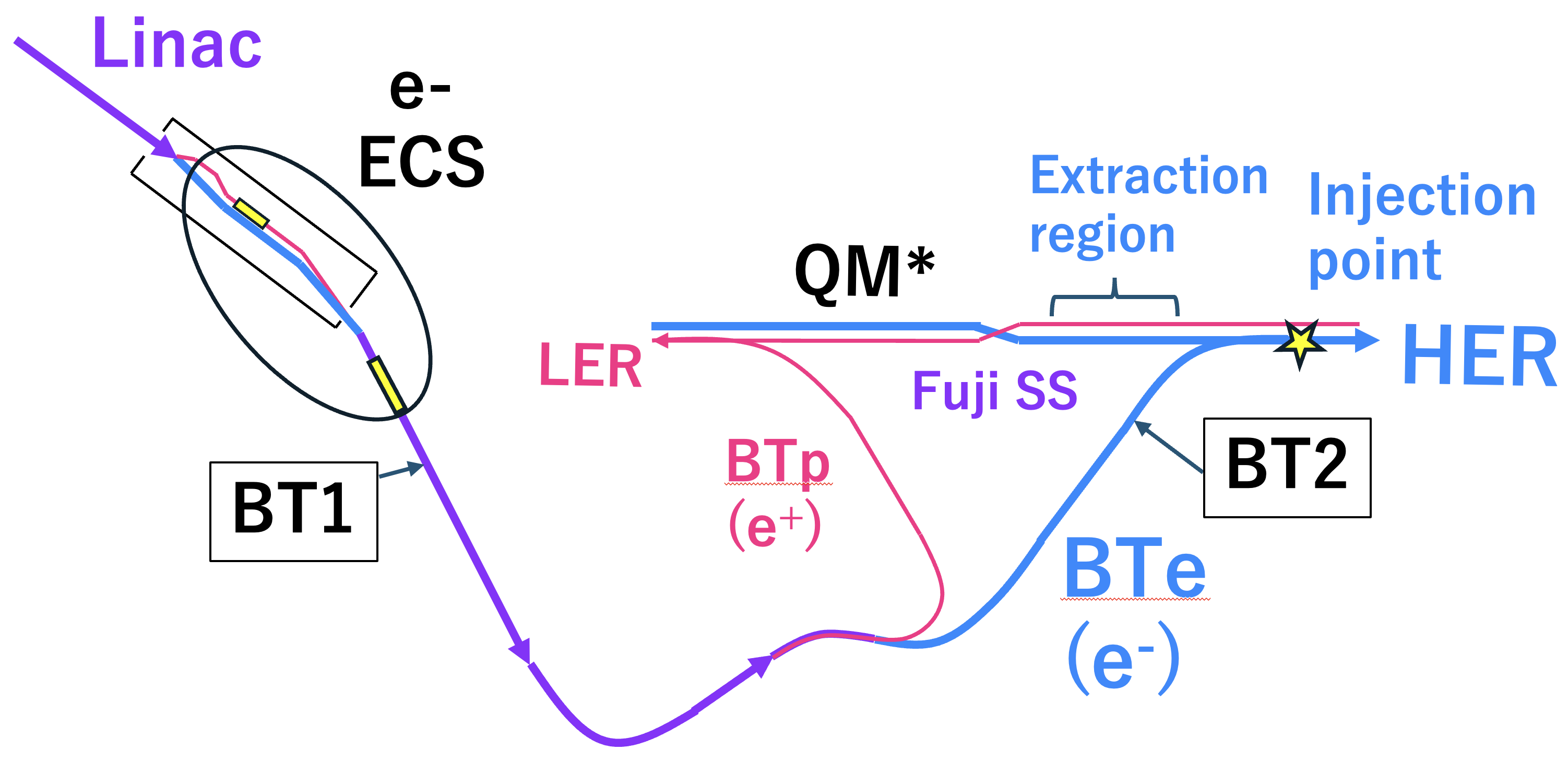}
   \caption{Layout of the injection region for SuperKEKB.
   The electron beam from the end of Linac ($s=0$\,m) is transported to the HER through an approximately 480\,m beam transport line (BTe) consisting of five arcs, including an energy compression system (ECS). The emittance is typically measured at BT1 ($s=174.5$\,m) and BT2 ($s=424.5$\,m) using beam profile monitors. SuperKEKB has four straight sections; the injection region is in “Fuji straight section'' (Fuji SS), and the quadrupole magnets used for tune adjustment are labeled {\tt QM*}. The beam extraction system is also installed closer to the injection point.
}
   \label{fig:layout}
\end{figure}

The injection efficiency to the HER has been examined by multi-particle tracking simulations through the electron BT (BTe) and the HER. First the injected particles are prepared by tracking taking into account the longitudinal wakes through the S-band 7\,GeV Linac, incoherent quantum synchrotron radiation (ISR), and coherent synchrotron radiation (CSR) through BTe. No machine errors are assumed here. The tracking was done in 6D using {\tt SAD}\cite{SAD} for Linac and {\tt elegant}\cite{ELEGANT} for BTe, respectively. The latter evaluates the CSR by a parallel plate model. The newly installed energy-compression system (ECS) near BT1 is also included. Those locations are shown in Fig.~\ref{fig:layout}.
The beam emittances from Linac are normalized to the measured values, by wire scanners, at BT1 in BTe.  In the actual beam operation, the emittances measured by an optical transition radiation (OTR) screen at BT2 were significantly larger than those at BT1\cite{eeFACT2025}, which have not been explained.
%by the tracking. 
Therefore we have artificially re-normalized the transverse emittances of the tracked particles just before the injection  to the measured values $\gamma\varepsilon_{x/y}$=200/150\,{\unit \um}, which are typical measured numbers at BT2. 
%Also the injected particles' transverse phase space (Fig.~\ref{fig:action}) is matched to the Twiss parameters as the of the HER at the injection point, having the design momentum offset $\delta_{\rm inj}$ for SI. 
Also the injected particles' transverse phase space (Fig.~\ref{fig:action}) is matched to the Twiss parameters optimized to minimize the action of the injected beam in the HER, 
having the design momentum offset $\delta_{\rm inj}$ for SI. 
%In the case of SI, the ring Twiss parameters are obtained for the off-momentum optics with $\delta_{\rm inj}$.
Such a matching in the real beam operation is considered to be achieved by tweaking several quadrupoles at the end of BTe by maximizing the injection efficiency to the HER.

\subsection{Injection Efficiency}
The tracking in the HER has been done by {\tt SAD} in 6D, which includes the full lattice with the crab-waist as well as known multipole errors in the final quadrupoles (QCS). 
The collimations have not included.
The ISR and an optional weak-strong beam-beam effects at the IP are taken into account. 
No unknown machine errors are included, relying on the optics correction of the HER, which routinely obtains about 5\,\% beta-beats in both planes, as well as dispersion and $x$-$y$ coupling corrections. 
The tracking has been performed up to 8000 turns after the injection.

\begin{figure}[htb!]
   \centering
   \includegraphics*[width=1\columnwidth]{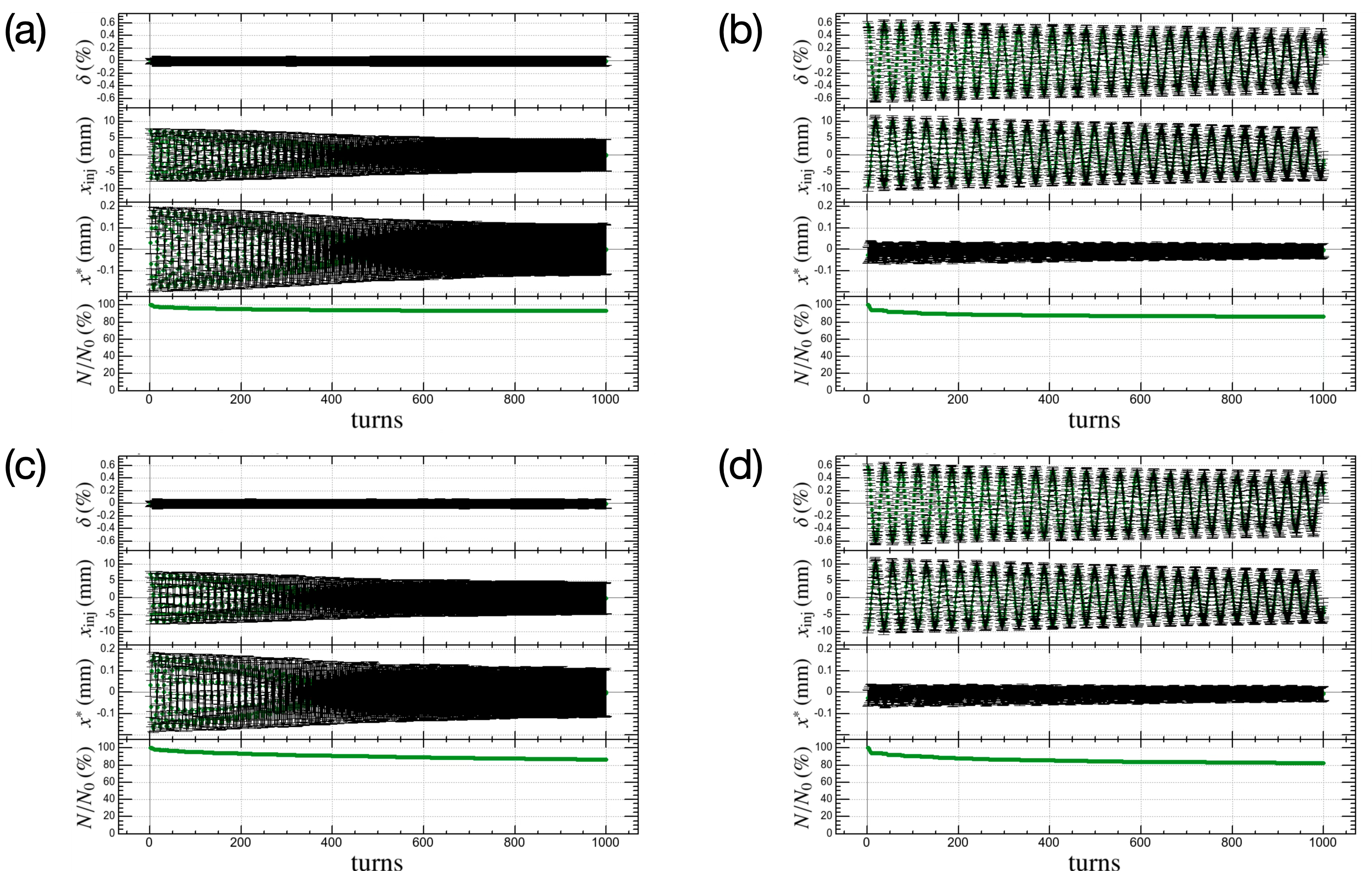}
   \caption{Behaviors of the injected beam for 1000 turns, (a,c): betatron and (b,d): synchrotron injections. 
   A beam-beam effect is turned on for (c,d) with $N_{e^+}=9.04\times10^{10}$.
   The rows show energy offset, horizontal positions $x_{\rm inj}$/$x^*$ at the injection point/IP, and the ratio of survived particles, respectively. The ring optics are for $\beta^*_{x,y}=(60,1)\,{\rm mm}$. 
    The fractional part of the tunes are set at $\{\nu_{x,y}\}=(0.531,0.575)$ for all cases. The synchrotron tune $\nu_z$ is -0.027 for (a,c), and -0.028 for (b,d), respectively. The number of injected particles was 4000.
    %, and the survival ratio are shown in Table~\ref{tab:results}.
}
   \label{fig:injeff}
\end{figure}

Fig.~\ref{fig:injeff} shows an example of the injection decays for BI and SI, without/with beam-beam.
Actually as we did not test SBI in the commissioning this time, we hereafter concentrate on the simulations only on BI and SI.  
The injection efficiency was around 90\,\% and 85\,\% for the BI and SI, respectively. 
In the results of SI, a beam loss by $\sim5\,\%$ at the first a few turns is noticeable, and it makes the injection efficiency slightly worse than BI. 
The reason has not been perfectly identified, but a possible reason is that the momentum acceptance of the HER is slightly smaller than the required. 
If we reduce the injection energy by 0.05\,\%, the initial beam loss disappears. 

It is difficult to express about their differences only from these results, so we tried a scan over the betatron tunes in the next subsection. 
The parameters for the positron strong beam is set to those roughly corresponding to the luminosity ${\cal L}=5\times10^{34}/{\rm cm}^2s$, with $\varepsilon_{x,y}=(4.0,0.04)\,{\rm nm}, \beta^*_{x,y}=(60,1)\,{\rm mm}$, except for the bunch population($N_{e^+}$), which has been increased from 4.34$\times10^{10}$ to 9.04$\times10^{10}$ in these simulations.
In Fig.~\ref{fig:injeff}, the horizontal amplitude $x_{\rm inj}$, in the second column looks smaller for the BI (a,c) than for the SI (b,d). 
This was because the separation between the stored and injected beams are smaller for the BI, as shown in Fig.~\ref{fig:action}. 
However, this horizontal displacement at the injection point mostly consists of the synchrotron motion in the case of SI. 
On the other hand, the horizontal oscillation at the IP, $x^*$, shown in the third row of Fig.~\ref{fig:injeff}, reaches an amplitude of approximately $\pm$0.2\,mm in BI, whereas it is suppressed to below $\pm$0.05\,mm in SI. 
This significant reduction in the horizontal injection oscillation suggests that SI provides improved stability during collision.

\subsection{Sextupole Misalignments}
To simulate the effect of residual $x$-$y$ coupling in the ring on beam injection, intentional vertical misalignments of the sextupole magnets were introduced in the simulation. 
The injection efficiency (survival ratio) was investigated for both betatron injection (BI) and synchrotron injection (SI) in the presence of sextupole misalignments. 
In both cases, the vertical misalignments of all sextupole magnets were generated using Gaussian random numbers. 
First, the rms of the distribution was set to 0.3\,mm, and values exceeding three sigmas were excluded. 
In the simulations, 12 different random seeds were used, and the results are shown in Table~\ref{tab:results}. %together with the results obtained without sextupole misalignments. 
\begin{table}[!hbt]
    \centering  
    \caption{Survived ratio (\%) at 8000\,turns after injection. Notations: bb=beam-beam($N_{e^+}=4.3 \times 10^{10}$), and $\Delta_{\rm sext}$=sextupoles' misalignments of 0.3\,mm(rms), with 1000 particles.}
    \begin{tabular}{lccc}
    \toprule
  & \textbf{BI} & \textbf{SI} & \textbf{SBI} \\
\midrule
w/o bb & 93.6$\pm$0.8 & 85.6$\pm$1.0 & $77.6\pm1.3$ \\
w bb  & 91.2$\pm$0.8 & 84.6$\pm$1.1 & $70.9\pm1.5$ \\
w bb \& $\Delta_{\rm sext}$  & 86.2$\pm$1.6 & 83.4$\pm$0.5 &  \\
%w/o bb & 93.2$\pm$0.35 & 84.9$\pm$0.53 & 75.3$\pm$0.67 \\
%w bb  & 86.4$\pm$0.50 & 80.4$\pm$0.60 &  \\
\bottomrule
   \end{tabular}
   \label{tab:results}
\end{table}
%The results are summarized in Table~\ref{tab:results}.
In Table~\ref{tab:results}, the size of error bars is derived by assuming the beam losses are completely random, which may not be applicable to SI, where some losses are concentrated within the first a few turns. %The decrement of the survival ratio is degraded by beam-beam, stronger for BI than SI.
The emittance ratio (vertical to horizontal) was $0.28\pm0.07\,\%$ and $0.25\pm0.05\,\%$ for the BI and SI lattices, respectively. 
A typical emittance ratio after optics corrections in real operation is approximately 1\,\%.  
Therefore, the $x$-$y$ coupling in these simulations appears to be conservative, and simulations with larger misalignment sizes of 0.4 and 0.5 mm were also performed. 
\begin{figure}[htb!]
   \centering
   \includegraphics*[width=1\columnwidth]{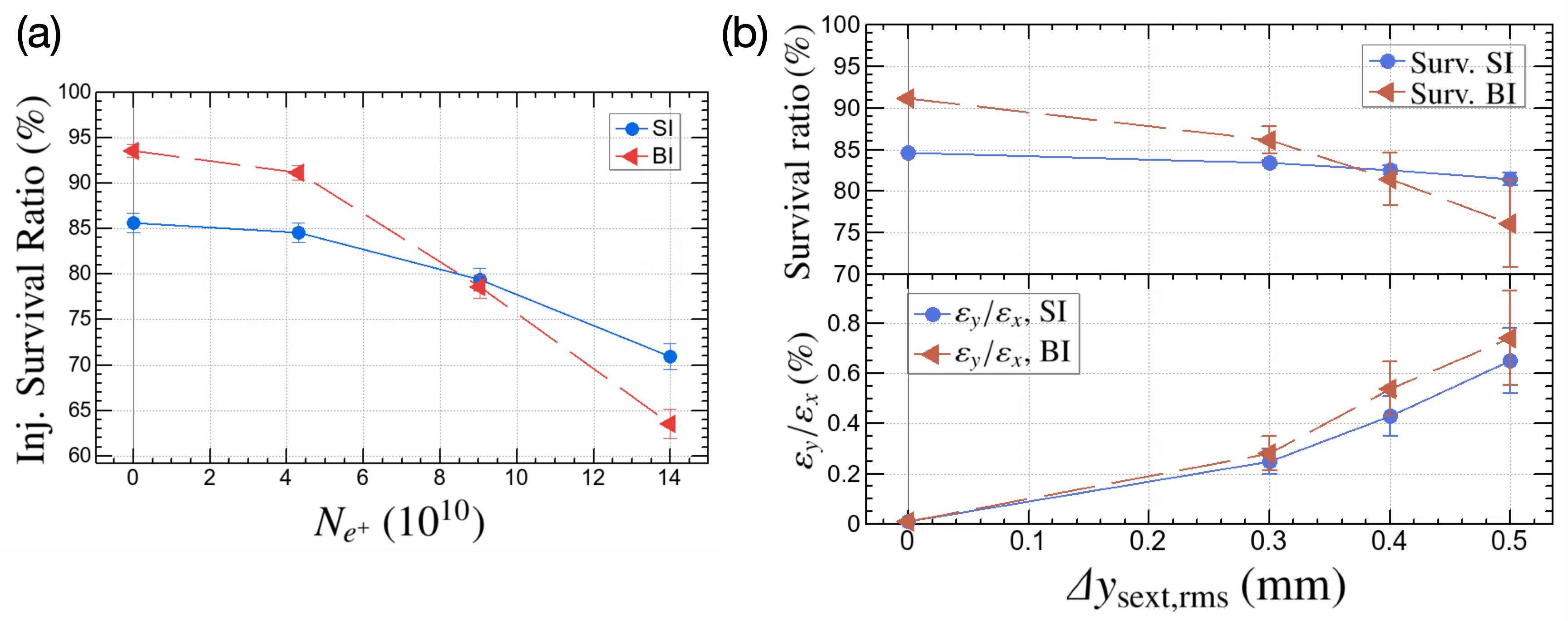}
   \caption{(a): The simulated dependence of the injection survival ratio on the bunch population of positrons,  obtained by tracking with 1000 particles after 8000 turns from injection. The intensity $N_{e^+}=4.3\times10^{10}$ corresponds to the highest luminosity record $5.24\times10^{34}/{\rm cm}^2{\rm s}$. (b): The simulated dependence of the survival ratio (upper) and the vertical/horizontal emittance ratio (lower) by random vertical misalignments for all sextupoles($\Delta y_{\rm sext}$) in the HER, with $N_{e^+}=4.3\times10^{10}$.}
   \label{fig:bunchinten}
\end{figure}
The dependence on $N_{e^+}$ and the sextupole misalignments are simulated as shown in Fig.~\ref{fig:bunchinten} (a) and (b), respectively.
From Fig.~\ref{fig:bunchinten}-(a), the stronger beam-beam effect will be affect to the injection for BI than SI.
From Fig.~\ref{fig:bunchinten}-(b), the effect of $x$-$y$ coupling on beam injection is more pronounced in BI. 
This result is to be expected, as BI involves larger horizontal betatron oscillations, which may influence beam injection via the $x$-$y$ coupling, touching the vertical dynamic aperture.
%Fig.~\ref{fig:bunchinten} shows the dependence of the survival ratio of the injected beam on beam–beam interaction or sextupole misalignments.
%As is clear from Fig.~\ref{fig:bunchinten}, 
BI is much more strongly affected by beam–beam interaction and sextupole misalignments than SI.
This can be a motivation for SI for higher luminosities. 

\subsection{Dependence on Betatron Tunes}

\begin{figure}[htb!]
   \centering
   \includegraphics*[width=1\columnwidth]{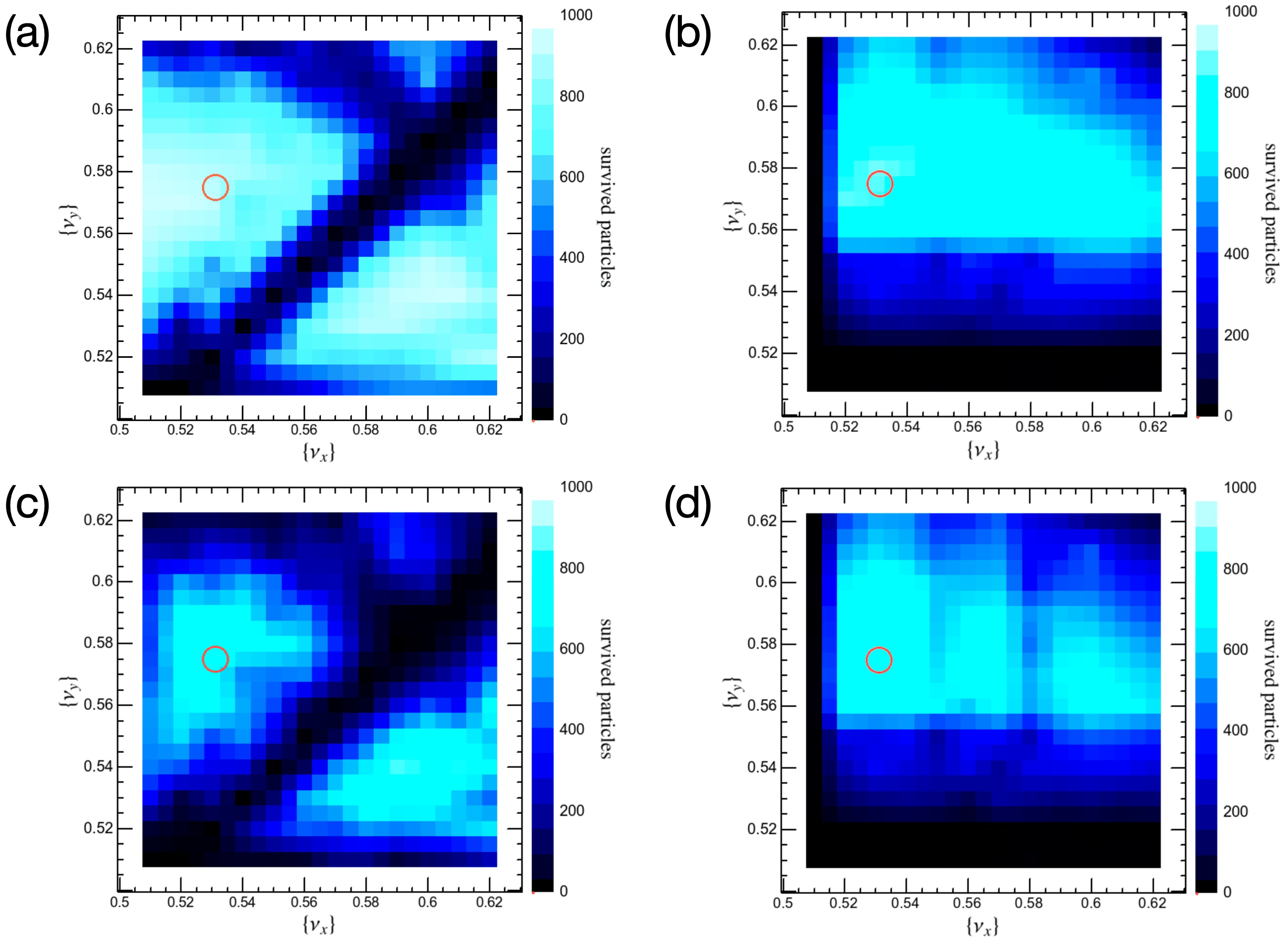}
   \caption{The dependence of the injection efficiency on betatron tunes. (a,c): BI, (b,d): synchrotron injection, and (a,b): no beam-beam, (c,d): beam-beam with $N_{e^+}=9.04\times 10^{10}$. Tracked 1000 particles for 8000 turns at each tune. The sextupoles are optimized for the DA at the tune shown by the red circle, and kept unchanged during the scan.}
   \label{fig:tunescan}
\end{figure}

We have examined the betatron tune dependence of the injection efficiency, by scanning the fractional tunes $\{\nu_{x,y}\}$. 
The results are shown in Fig.~\ref{fig:tunescan}. The tunes are scanned by changing 6 quadrupoles {\tt QM*} in a dispersion-free area in the Fuji SS. 
The number of quadrupoles is just necessary and sufficient to change the tunes while maintaining the optics of the rest of the ring. 
The sextupoles are once optimized for the dynamic aperture at one tune, and kept unchanged during the scan.

A few things are noticed in this scan:
(1) the differences due to beam-beam are visible in both cases, and stronger in BI. 
(2) Clear resonance lines $\nu_x-\nu_y\approx N$ and $5(\nu_x+\nu_y)\approx N$ are seen only for the BI, which are strengthened by beam-beam. 
(3) there are strong dark bands if the tunes get close to $\{\nu_{x,y}\}\rightarrow0.5$ for SI. The reason for the dark bands is that the optics does not have sufficient transverse momentum acceptance required by SI for $\{\nu_y\}\lesssim0.55$, as examined later in subsection \ref{ss:DAXY}.

\subsection{Vertical emittance of injected beam}

In the simulations described above, the normalized vertical emittance ($\gamma \varepsilon_y$) of the injected beam was assumed to be 150~$\mu\mathrm{m}$, which is a typical value during routine beam operation. However, this parameter can vary depending on the beam-tuning conditions in the linac and the beam transport line. To investigate the dependence of the injection efficiency on the vertical emittance, simulations were performed with artificially varied values of the injected-beam vertical emittance.

\begin{figure}[htb!]
   \centering
   \includegraphics*[width=0.6\columnwidth]{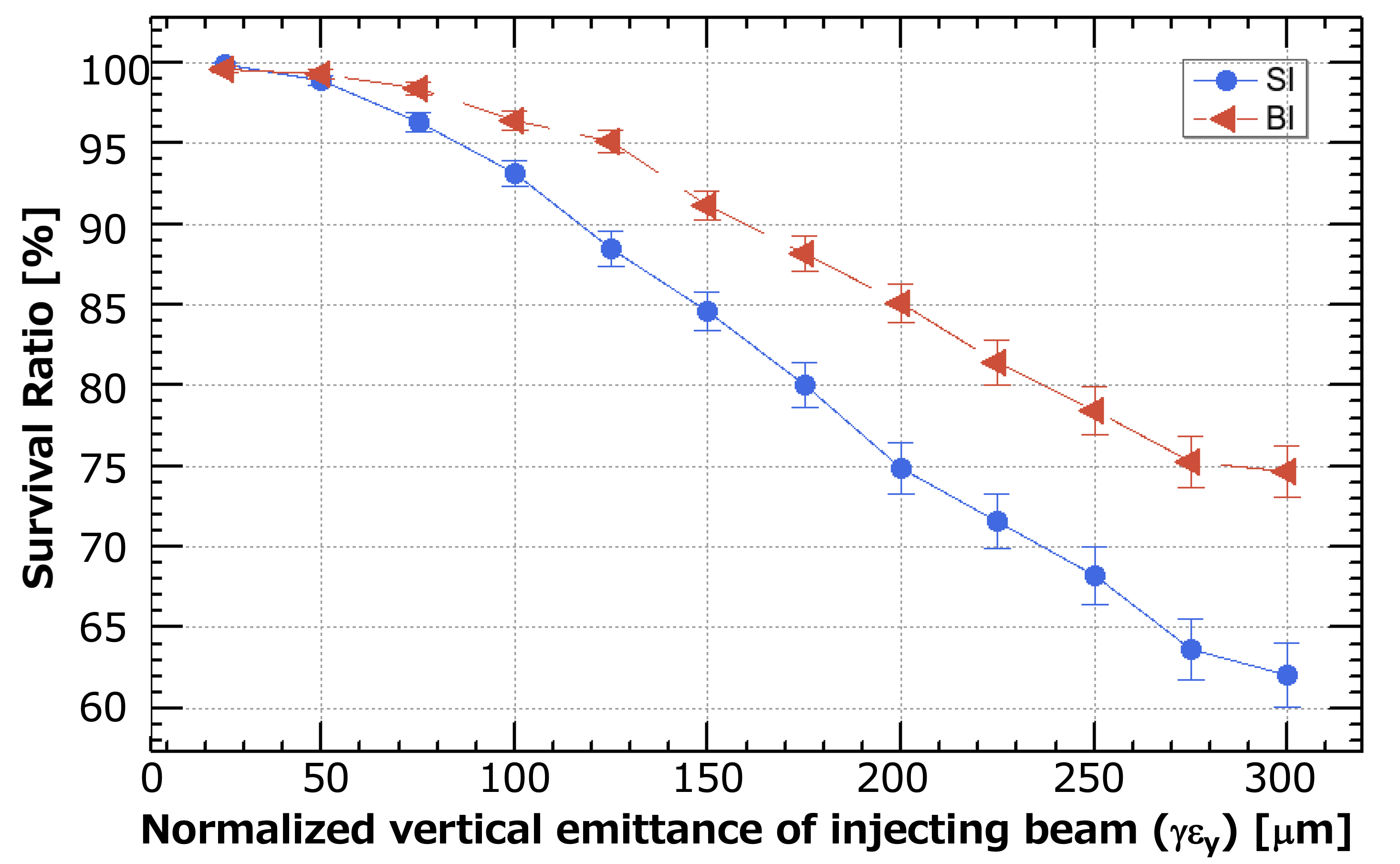}
   \caption{Simulated dependence of the injection survival ratio on the normalized vertical emittance of the injected beam. The results were obtained by tracking 1000 particles for 8000 turns after injection. A positron bunch population of ($N_{e^+}=4.3 \times 10^{10}$) was assumed, and no vertical misalignments were applied to the sextupole magnets. All other simulation parameters were identical to those used in the simulations described above.}
   \label{fig:verticalemittance}
\end{figure}

In these simulations, the collimator apertures were not included. This simplification was justified by a separate study with ($\gamma \varepsilon_y = 150~\mu\mathrm{m}$), which showed no significant difference in the injection efficiency between simulations with the horizontal and vertical collimator apertures set to their values used in physics operation and simulations without collimators. This result indicates that the injection efficiency is limited by the dynamic aperture rather than by the physical aperture.

 Figure~\ref{fig:verticalemittance} shows the simulated dependence of the injection efficiency on the normalized vertical emittance of the injected beam. The result indicates that one of the main causes of beam loss after injection is the limited vertical acceptance of HER relative to the large vertical emittance of the injected beam. The more rapid degradation of the injection efficiency with increasing vertical emittance in synchrotron injection can be attributed to the narrower transverse acceptance at larger energy offsets in synchrotron injection than in betatron injection.

 \subsection{Dynamic Aperture for the injected beam \label{ss:DAXY}}
\begin{figure}[htb!]
   \centering
   \includegraphics*[width=0.9\columnwidth]{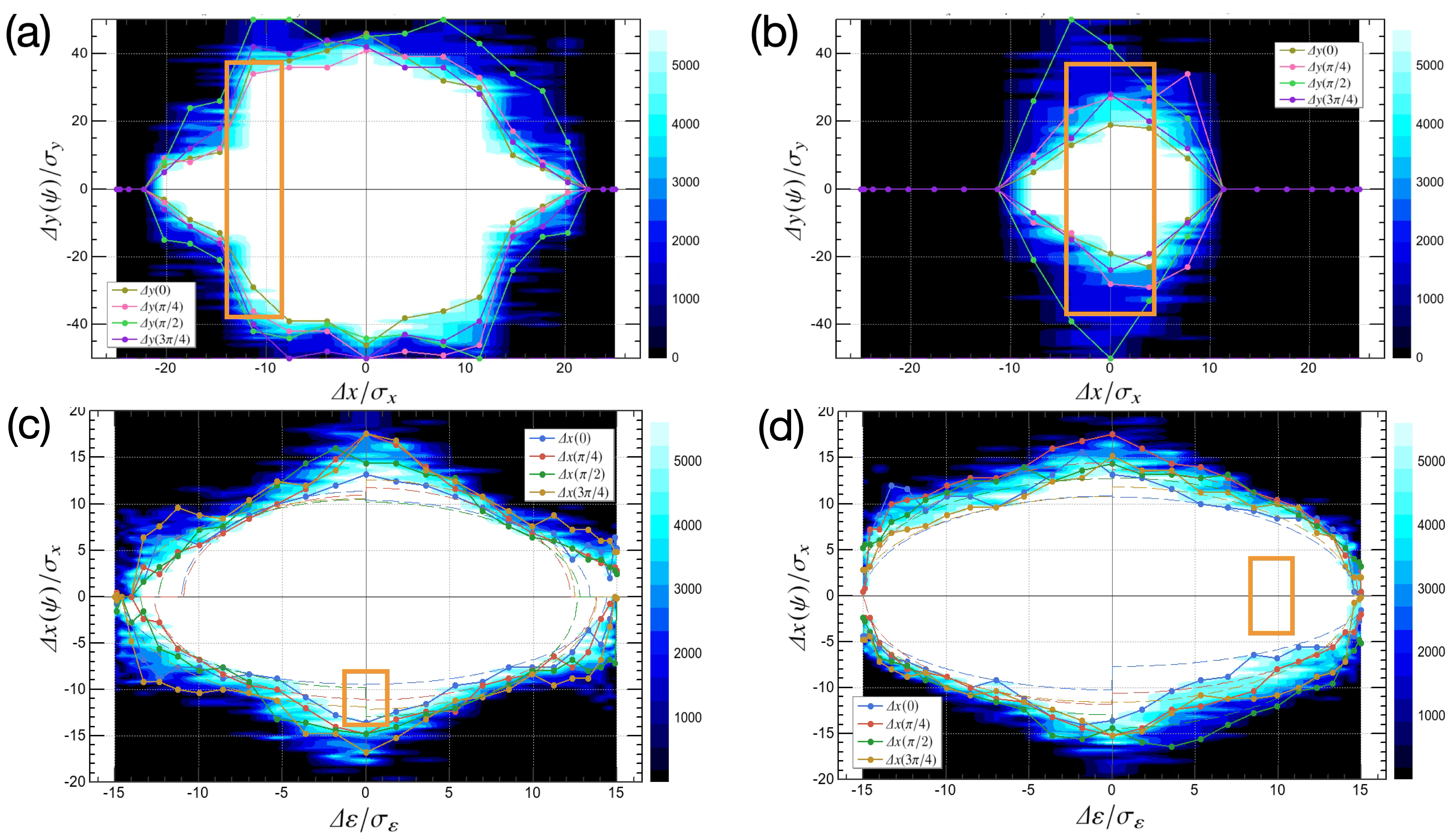}
   \caption{The transverse dynamic apertures for (a) BI and (b) SI. Plotted for the initial displacements $\varDelta x$ and $\varDelta y$. The orange square indicates $\sim\pm2.5\sigma$ of the injected beam($\gamma\varepsilon_{x/y}$=200/150\,{\unit \um}) on each plot. As for SI, the tracking starts with an  initial momentum offset $\varDelta p/p_0=0.6$\,\%, the perform a synchrotron motion. The dynamic momentum acceptance are plotted in (c) BI and (d) SI, respectively, where the horizontal axes are the initial momentum offset. The amplitude ratio $J_y/J_x$ are set at 1\,\%. The orange squares show the horizontal size and momentum spread of the injected beams. Note that he vertical sizes are ignored in plots (c) and (d). These DAs are obtained by tracking for 5800 turns, with $\pm4$ phases in $y$-$p_y$ plane (a,b), and both $x$-$p_x$ and $y$-$p_y$ planes (c,d), while the initial $\varDelta p_x=0$ in (a,b). No quantum effect of the synchrotron radiation is applied. Beam-beam effects are included with $N=4.3\times10^{10}$.
   }
   \label{fig:injDA}
\end{figure}

The dependence on the vertical emittance of the injected beam above becomes more understandable if we look at the dynamic aperture (DA) for the injected beam. Figure\,\ref{fig:injDA}(a,b) plots the transverse dynamic apertures for the injected beam for BI and SI. As for BI, the injected beam is shifted in the $x$-direction given by the gap by the injection septum as shown in Fig.\ref{fig:injDA}(a). On the other hand, for SI, the DA in Fig.\ref{fig:injDA}(b) is obtained by giving the momentum offset $\varDelta p/p_0\approx 0.6\,\%.$ The survival ratios of the injected beam with $\gamma \varepsilon_y = 150~\mu\mathrm{m}$ are roughly consistent with the previous results in Fig.~\ref{fig:verticalemittance}.

According to the DA, SI may have more strict limitation on the vertical injected emittance than BI, due to the narrowness of the off-momentum DA. This situation depends on the ring lattice as well as the emittance of the injected beam.

\subsection{Photons Hitting the IP Beam Pipe}
The number of primary photons from the injected beam was estimated by tracking for the first 30 turns with 20000 particles. The IP beam pipe is located at $\pm0.1$\,m from the IP, and has 10\,mm inner radius, tilted to the $e^+$ beam by the half crossing angle 41.5\,mrad. For the BI, the number of photons per 1000 electrons per turn and average photon energies were $1.65\pm0.045$ and 0.37\,keV, respectively. For SI, those were $0.69\pm0.029$ and 0.39\,keV.

\section{Injection Commissioning}
During the startup of SuperKEKB in autumn 2025, the SI was tested from November 6 to 20. Extensive vacuum works, particularly in the positron low-energy ring (LER) during the preceding summer, significant beam conditioning time was needed. In parallel, vacuum conditioning and preparations for SI were carried out in the HER.
Furthermore, also during the Linac summer shutdown, the RF gun for the electron beam was replaced, and beam tunings were needed to achieve low emittance levels for the beam.
The operation was performed with $\beta^*_{x,y}$ of (400, 80.9)\,mm (so called ``detuned optics''), (200,8)\,mm, (100,3)\,mm, and (60,1)\,mm for four days, two days, two hours, and 17 hours, respectively.
Finally, stable collisions for physics data taking were achieved.
\begin{figure}[!htb]
   \centering
   \includegraphics*[width=1\columnwidth]{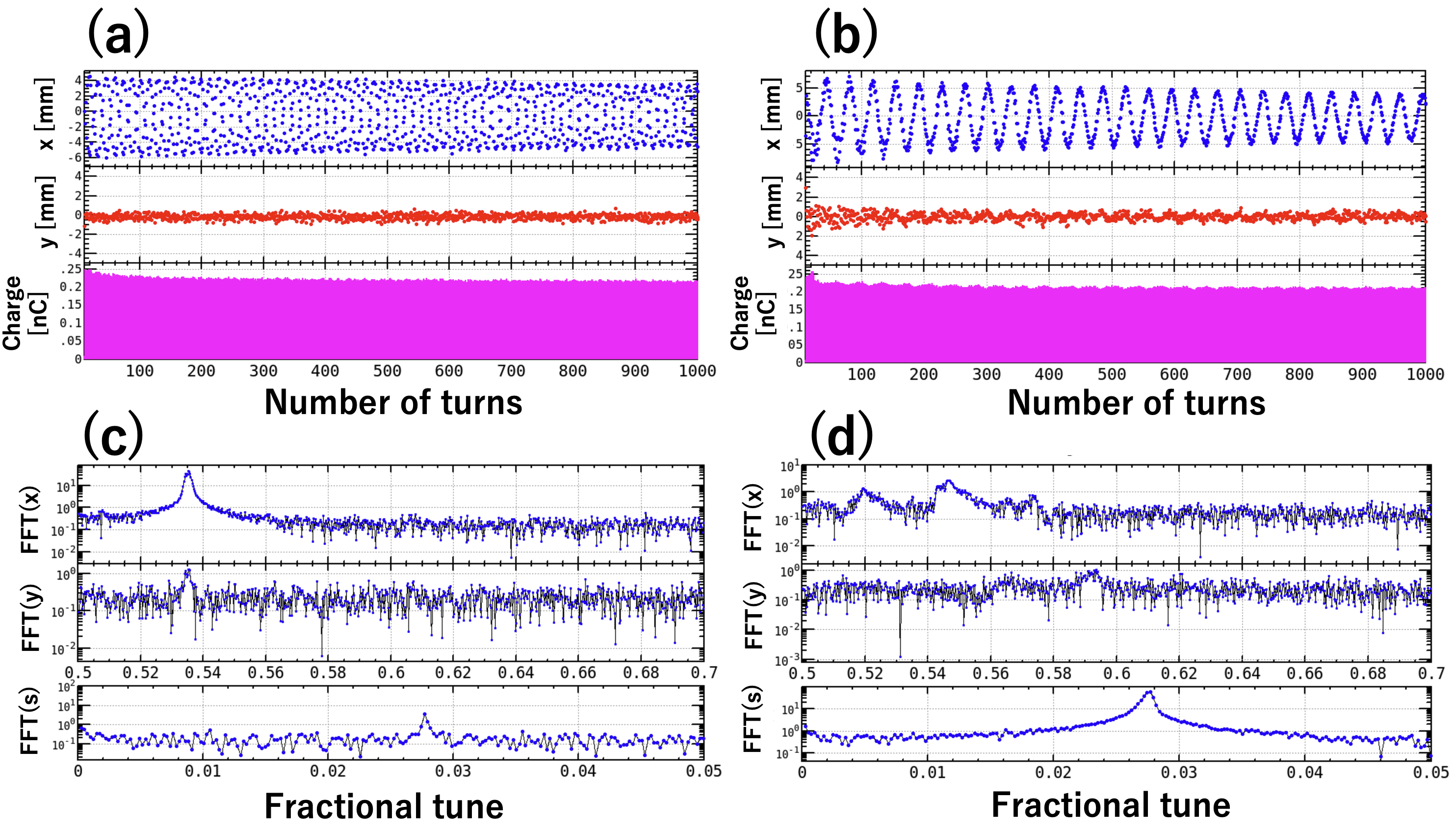}
   \caption{Measured beam oscillations using TbT-BPMs with (a,c) BI and (b,d) SI. Top: horizontal / vertical positions and beam charge (up to 1000\,turns). Bottom: amplitudes of horizontal, vertical, and longitudinal frequency components.}
   \label{fig:TbT}
\end{figure}

During the detuned optics, the oscillations of injected beam were measured with the TbT-BPMs, as shown in Fig.~\ref{fig:TbT}.
A significant difference in oscillation frequencies is observed between (a) BI and (b) SI. 
A peak appears in the frequency spectra at the horizontal tune in (c), while the synchrotron tune in (d).
These results indicate a perfect realization of the synchrotron injection as the simulation results shown in Fig.~\ref{fig:injeff}.
The loss in the first few turns, as mentioned in Fig.~\ref{fig:injeff}, has been indeed actually observed in Fig.~\ref{fig:TbT}(b).
%When the HER optics was detuned or at $\beta_y^*$=8\,mm, up to ~10\% of the nominal current was occasionally observed in an adjacent bucket, suggesting unintended injection. 
%This was not reproducible in tracking simulations, even with an enlarged longitudinal distribution (×XXX), so it is unclear whether the effect is specific to synchrotron injection. 
%As no logs are available, future measurements are needed to confirm whether simultaneous injection into neighboring buckets occurs.

\begin{figure}[!htb]
   \centering
   \includegraphics[width=0.6\textwidth]{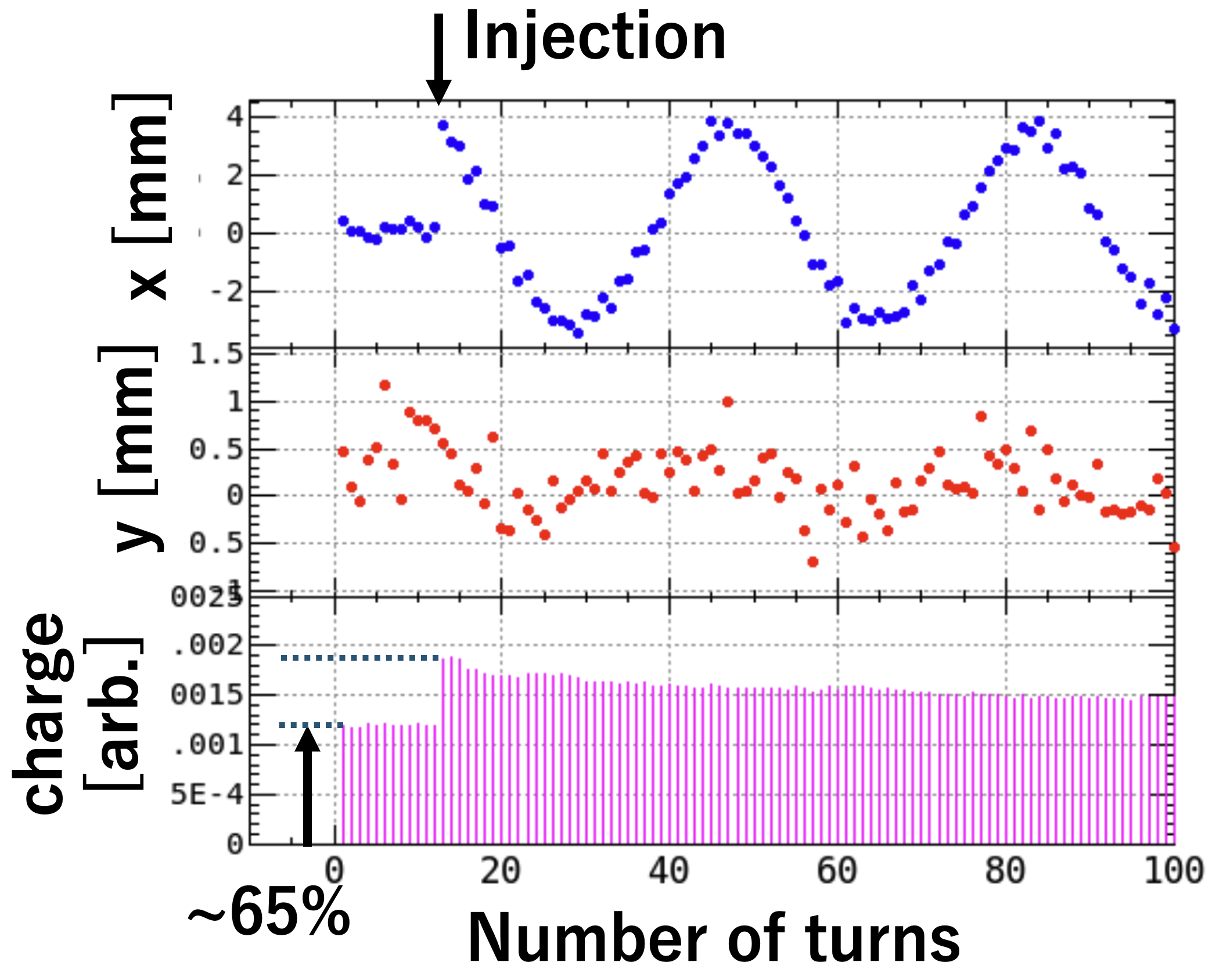}
   \caption{A TbT-BPM measurement of the injected beam. 
The horizontal axis shows the turn number, where the injection was performed at the turn \#12.  The vertical axes (top to bottom) show horizontal position, vertical position, and beam charge. The points before the turn \#12 show the previously stored beam, which was kicked out by the injection kickers at the turn \#12, while the next beam is injected simultaneously. The stepwise increase in beam charge indicates that the survival ratio of the previously injected beam was 65\,\%, by assuming the same amount of injected charge.}
\label{fig:TbT-inj}
\end{figure}

At $\beta_y^*$=1\,mm, the HER dynamic aperture is small, making injection tuning difficult in both injection scheme. 
Vertical collimators, tunes, and injection orbits were adjusted as needed, while the horizontal collimators were kept almost fully open with little tuning for SI, whereas for BI, they should be tuned more carefully. 
The RF phase setting between the injected beam and the ring is the key difference between SI and BI.
In SI, since the injected beam energy differs from that of the ring, the longitudinal oscillation cannot be reduced to zero. 
Instead, as shown in Fig.~\ref{fig:TbT-inj}, the phase was adjusted so that the horizontal oscillation at a TbT-BPM in a dispersive region is maximized on the first turn. 
This ensures that the longitudinal oscillation arises solely from the energy difference, {\it i.e.}, the phase is properly tuned.
\begin{figure}[!htb]
   \centering
   \includegraphics*[width=1\columnwidth]{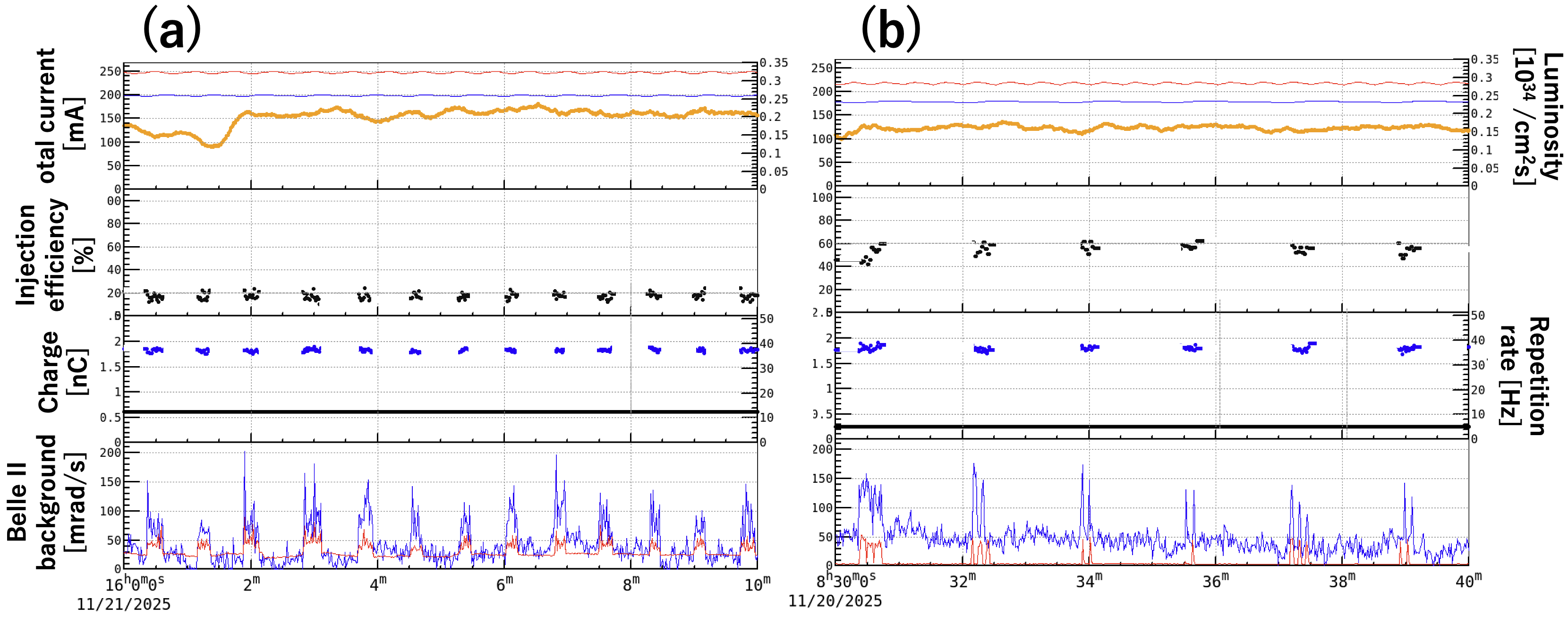}
   \caption{Ten-minute collision operation for (a) BI and (b) SI at $\beta_y^*$=1\,mm; (a) was taken 1.5 days after the SI study. 
   Top: stored currents (red: LER, blue: HER) and luminosity (yellow). 
   Second: injection efficiency (black). 
   Third: injected charge (blue) and repetition rate (black). 
   Bottom: two types of background in the Belle II detector.}
   \label{fig:trend}
\end{figure}

The collisions started 7 hours after optics tuning of the HER to $\beta_y^*$=1\,mm, followed by 7.5 hours of physics data taking. 
Fig.~\ref{fig:trend} compares BI and SI under similar stored currents and injected charge. 
The injection efficiency with SI was about three times higher than with BI, while the background in the Belle II detector appears comparable.
Since both use top-up injection, the injection is not continuous; periods with large background in the Belle II detector correspond to injection timing. During BI, the frequency of injection was higher because the HER beam lifetime happened to be shorter. Note that the ring optics for BI used for this beam operation was not the optimized one used in the simulation in the previous section. 
With SI, an injection efficiency of 60\,\% was achieved within 7.5 hours of tuning including collimators, whereas BI did not reach the same efficiency even after four days. This was just a single example, and it was not clear why BI's startup took so long this time, but we may say that at least the injection tuning for SI is not an issue.
This indicates that SI is considerably easier to optimize owing to the absence of horizontal oscillations.

%However, shortly afterward, when the injection repetition rate was increased from 5\,Hz to 12.5\,Hz, injection became unstable and frequently stalled, forcing the termination of the SI test period.
%Since this phenomenon was later observed occasionally even during BI, it is likely not specific to SI but rather related to the higher repetition rate.

\section{Conclusion}
Synchrotron injection for colliding \& top-up was tested for the first time in the HER of SuperKEKB. 
Simulations indicated that SI is more insensitive than BI on beam-beam effects, sextupole misalignments, and the synchrotron light hitting on the IP beam pipe. However, simulations reveal that SI may be suffered by the narrowness of DA for off-momentum particles, especially when the emittances from the injector is large.

In the experiment, the injection oscillation was fully converted from the horizontal to the longitudinal planes, demonstrating a successful SI operation.

At $\beta_y^*$=1\,mm, tuning was simpler than for BI, and an injection efficiency of ~60\,\% was achieved, which was not worse than the regular operation of BI.
\begin{acknowledgments}
The authors thank the all SuperKEKB members of Linac/Ring/Belle II and MELSC operators for performing the beam operation. 
They also thank A. Faus-Golfe, P. Bambade, and F. Zimmermann for giving chances for presentations and discussions. 
\end{acknowledgments}


\end{document}